\documentclass{elsart}

\makeatletter
\def\ps@copyright{\let\@mkboth\@gobbletwo
  \def\@oddhead{}%
  \let\@evenhead\@oddhead
  \def\@oddfoot{\small\slshape 
   TPR-99-02\hfil\@date\/}%
  \let\@evenfoot\@oddfoot
}
\makeatother

\usepackage{epsfig}
\newcommand{\bra}[1]{\big< #1 \big|}
\newcommand{\ket}[1]{\big| #1 \big>}
\newcommand{\slas}[1]{\FMslash{#1}}
\newcommand{\Slas}[1]{\FMSlash{#1}}
\newcommand{\CPC}[3]{Comput.\ Phys.\ Commun.\ {\bf {#1}}, {#2} ({#3})}
\newcommand{\PRT}[3]{Phys.\ Rep.\ {\bf {#1}}, {#2} ({#3})} 
\newcommand{\PRD}[3]{Phys.\ Rev.\ D {\bf {#1}}, {#2} ({#3})}
\newcommand{\PRL}[3]{Phys.\ Rev.\ Lett.\ {\bf {#1}}, {#2} ({#3})}
\newcommand{\NPB}[3]{Nucl.\ Phys.\ B {\bf {#1}}, {#2} ({#3})}
\newcommand{\PLB}[3]{Phys.\ Lett.\ B {\bf {#1}}, {#2} ({#3})}
\newcommand{\ZPC}[3]{Z.\ Phys.\ C {\bf {#1}}, {#2} ({#3})}
\newcommand{\EPJC}[3]{Eur.\ Phys.\ J.\ C {\bf {#1}}, {#2} ({#3})}

\begin{document}
\begin{frontmatter}
\title{Diffractive charged meson pair production}

\author[R]{B.~Lehmann-Dronke},
\author[K]{M.~Maul},
\author[R]{S.~Schaefer},
\author[R]{E.~Stein},
\author[R]{A.~Sch\"afer}

\date{16 April 1998}

\address[R]{Institut f\"ur Theoretische Physik, Universit\"at
  Regensburg, D-93040 Regensburg, Germany}
\address[K]{Nordic Institute for Theoretical Physics, Blegdamsvej 17,
  DK-2100 Copenhagen, Denmark}

\begin{abstract}
We investigate the possibility to measure the nonforward gluon
distribution function by means of diffractively produced
$\pi^+\pi^-$ and $K^+K^-$ pairs in polarized lepton nucleon
scattering. The resulting cross sections are small and are
dominated by the gluonic contribution. We find relatively large spin
asymmetries, both for $\pi^+\pi^-$ and for $K^+K^-$ pairs.\\[0.3cm]
\end{abstract}

\end{frontmatter}

\section{Introduction}

QCD parton distribution functions
are specified by matrix-elements of nonlocal quark and gluon 
twist-2 operators separated by a light like distance and 
taken inside a forward nucleon matrix element.
Although these forward matrix elements were generalized long time
ago to nonforward parton distribution functions \cite{class},
where nonlocal QCD operators are sandwiched between states of unequal 
momenta, they became only recently the object of intense
studies \cite{news,Ra97,Bel98}.
This interest was driven by the hope that the measurement of 
nonforward parton distributions (\mbox{NFPDs}) and subsequent extrapolation
to the forward limit might shed more light on the spin and angular
momentum composition of the nucleon \cite{Ji97}.

Contrary to ordinary parton distribution functions, 
nonforward parton distribution functions are probed when the nucleon
recoils elastically. Thus \mbox{NFPDs} provide a natural interpolation
between ordinary parton distribution functions and elastic form factors. 
It was shown that \mbox{NFPDs} are in principle measurable in deeply virtual
Compton scattering or through exclusive production of light vector
mesons. 

\begin{figure}
\centerline{\epsfig{file=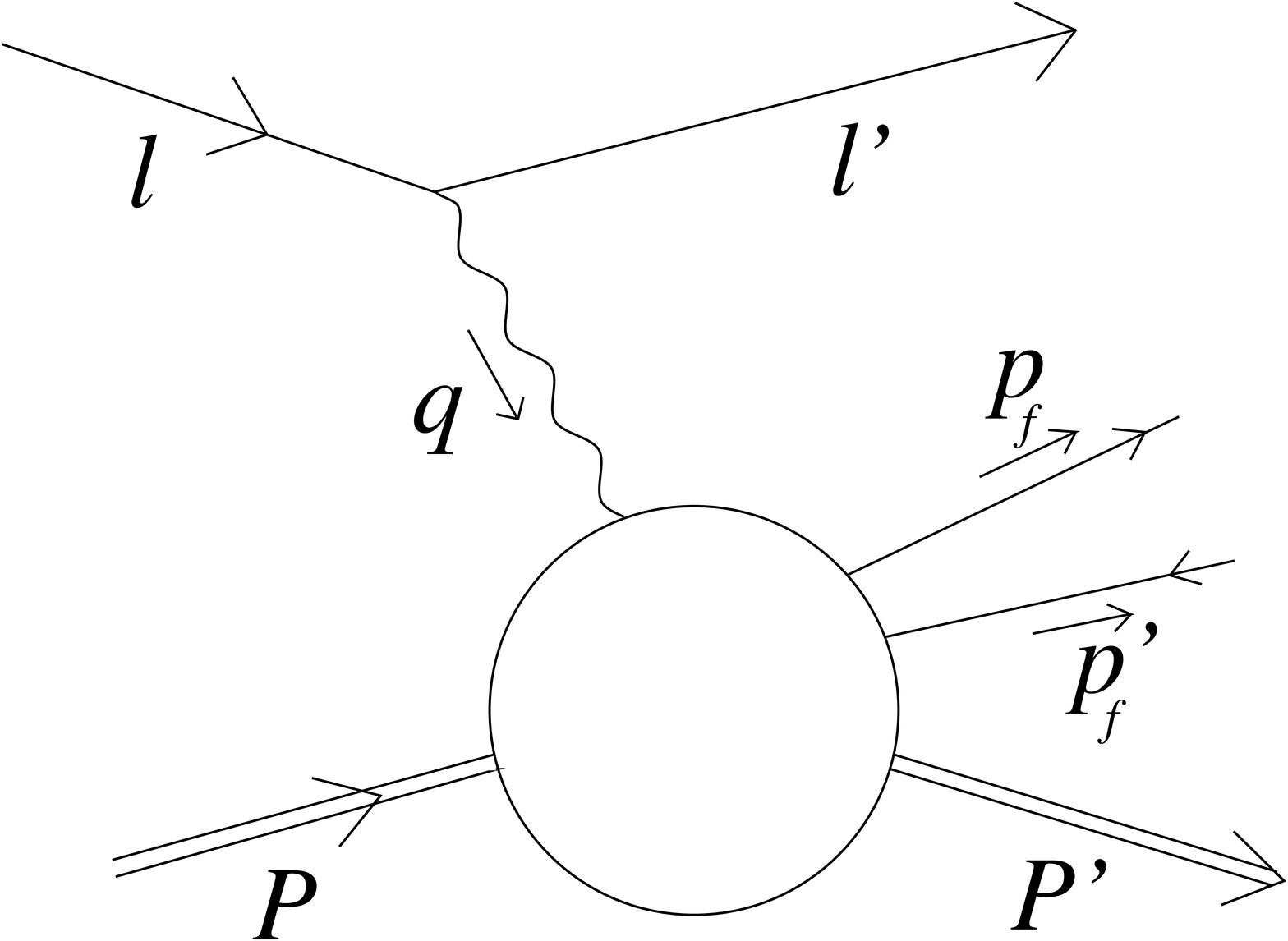,width=5cm}}
\caption{The process of the quark antiquark pair production to 
lowest order QED.}
\label{momenta}
\end{figure}

In this paper we shall consider the diffractive leptoproduction of 
quark antiquark pairs
in a kinematics where the produced quark pair does not form a single
meson but fragments into a $\pi^+\pi^-$ 
or $K^+K^-$ pair with comparatively large $p_T$ (Fig.~\ref{momenta}).
We will consider the case of
longitudinal polarized electron beam and proton target.
This process is quite similar to diffractive leptoproduction of dijets,
analyzed in \cite{Di95,GM98}, which was proposed to allow for a measurement
of the polarized glue in the nucleon.
We notice that at  medium energies like in the HERMES experiment dijets 
cannot be observed. In this experiment one can only look for single
hadrons representing the jets.
The idea to access the polarized forward gluon distribution function of the 
nucleon via the measurement of a double hadronic high momentum final state 
has been proposed in \cite{BHK98}.

We treat the fragmentation 
process of the two produced quarks into the final hadrons
using the LUND string fragmentation model 
\cite{strgfrg,Sj95}. To be precise we use this model to
extrapolate from dijet production, where it is known
to be valid, to the case where the two jets are represented by
e.g. $\pi^+$ and $\pi^-$. While we do not expect the LUND model to
give precise results in this few particle process it should allow for
an order of magnitude estimate.

In our calculation we will assume that two pion (kaon) 
leptoproduction can be factorized into three parts.
We have two soft parts: one describing the fragmentation of the two 
partons, the other the production of two 
partons from the elastic scattering off the nucleon.
The second soft part will be given by a nucleon matrix element of a
light cone operator parameterized by \mbox{NFPDs}.
The hard part connects the elastic scattering process and
subsequent fragmentation.

Technically, the process we consider is closely related to the 
exclusive production of a light vector meson. 
In such a case the proof of factorization has been given in
\cite{CFS97}. Recall that the proof  only holds for 
longitudinal photons. Meson production by a transversely  polarized photon 
is plagued by endpoint singularities as noted in \cite{BFGMS94}. 
Thus a general factorization theorem does not exist in this case.
Further the factorization theorem was proven only on the level of twist-2.
However, it was noted that the proof in \cite{CFS97} can be extended to 
two-pion production if these are produced close to threshold, i.e. if
the mass of the produced final state is much smaller than the virtuality
of the probing photon \cite{Diehl98}. 
In such a kinematics it is useful to describe the pion pair in terms
of a two-pion twist-2 distribution amplitude \cite{Pol98}. 
The main  difference to our calculation then is the treatment of the produced
meson in terms of a light cone twist-2 meson wave function rather than in 
terms of the fragmentation model. In our calculation we do not restrict
ourselves to the kinematic region $M^2(h^+ + h^-)<Q^2$ because we have
to avoid too small invariant masses in the final quark state to
justify the application of the LUND fragmentation model and for
correspondingly large $Q^2$ the counting rates are too small for an
experiment like HERMES. However, in our kinematics a hard scale is
given by $\mu^2=p_T^2(1+\frac{Q^2}{M^2(h^+ + h^-)})$ with $p_T$ being the
transverse momentum of the produced quark in the center of mass system
of the incoming nucleon and the virtual photon \cite{bar}. The hardness
of this scale ensures the applicability of perturbation theory.
The calculation therefore can be considered as a parton model description 
combined with the LUND fragmentation model.

\begin{figure}
\centerline{\epsfig{file=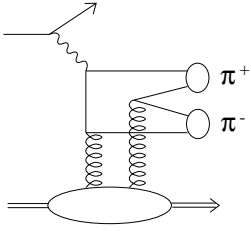,width=3.3cm}}
\caption{diagram for the production of a pion pair by the
fragmentation of four quarks.}
\label{strik}
\end{figure}

We would like to stress that the process we are considering 
is not the only one leading to two charged pions in the final state.
An exclusively produced $\varrho$-Meson for instance will immediately
decay into a pion-pair. However, the cuts in our 
numerical analysis make sure that we only count pions produced away
from the $\varrho$-resonance.
A pion pair can also be produced through the 
exclusive production of four quarks, each pair forming a meson%
\footnote{We are indebted to M.~Strikman for drawing our attention
to this point.}.
A leading order diagram for this process is shown in Fig.~\ref{strik}.
However, one may expect that such diagrams are kinematically suppressed
as can be seen best in the center of mass frame of the incoming
nucleon and the virtual photon: two of the four produced quarks are
obtained by the decay  of a gluon taken out of the nucleon and
therefore will carry a fraction of the incoming proton momentum while
the other quark antiquark pair carries the momentum of the virtual
photon. Without at least one additional internal gluon line, which
would lead to a suppression in $\alpha_S$, it is not possible to
rearrange the momenta such that two light mesons are formed.
Also we did not consider
the process of two gluon production and their fragmentation into a
meson pair. However, it turns out that in the considered kinematical
range contributions from gluon nonforward parton distributions
dominate compared to quark nonforward distributions whereas only the
latter lead to two gluon production in the leading order.

The paper is organized as follows: First, we define the 
general kinematics, give the formula for the partonic
cross section and derive the amplitudes for exclusive $q\bar{q}$
production. We then discuss  the models used for the polarized and
unpolarized \mbox{NFPDs}.
Subsequently, we discuss  the fragmentation process of
the two final partons into the meson pair using PYTHIA/JETSET
\cite{Sj95}.

\section{Cross Section for Diffractive $q \bar q$ Production}

\begin{figure}
\centerline{\psfig{figure=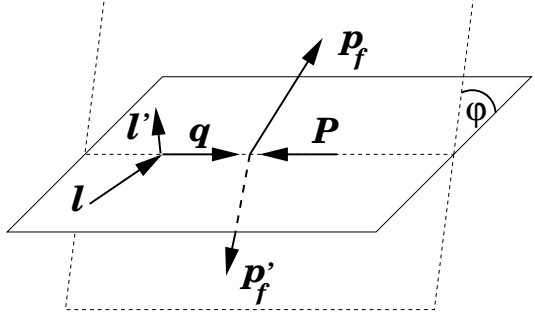,width=7cm}}
\caption{Kinematics in the boson-nucleon center of mass frame. }
\label{kinem}
\end{figure}

As already mentioned we assume that meson
pair production factorizes into hard quark pair production and
a fragmentation part. We treat the
fragmentation process on the level of transition probabilities using
the program JETSET. Therefore we have to calculate analytically only
the cross section for the process
\begin{equation}
e(l) + N(P) \to e(l') + q(p_f) + \bar{q}(p'_f) + N'(P') \;.
\end{equation}
The corresponding diagram to lowest order QED is shown in
Fig.~\ref{momenta}.
For the definition of the particle momenta see
also that figure. We work in the center of mass frame of the exchanged
photon and the initial nucleon and neglect nucleon and lepton masses.
Furthermore we take the momentum transfer to the nucleon to be
proportional to the initial nucleon momentum $P_\mu - P'_\mu = \zeta P_\mu$.
We have chosen the momenta $\vec{q}$ and $\vec{P}$ to define the $z$-
and $-z$-direction respectively as explained in \cite{PR80}. See also
Fig.~\ref{kinem}.

\begin{figure}[b]
\centerline{\psfig{figure=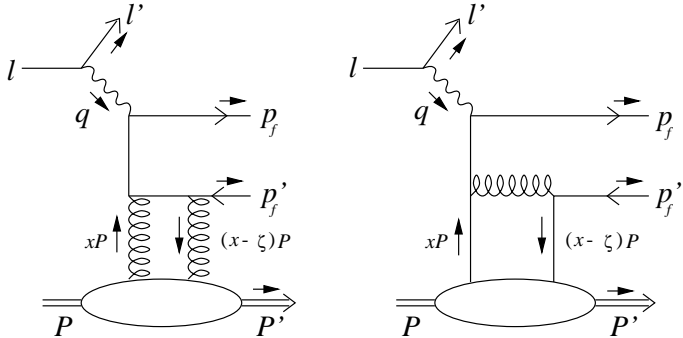,width=9cm}}
\caption{Typical diagrams to the process involving two gluon emission and two
quark emission of the scattered nucleon.} 
\label{dijet}
\end{figure}

The masses of the final quarks are neglected $p_f^2 = p_f'^2 = 0$.
Recall that for exclusive production of only one particle the on-shell
condition leads to the constraint that the skewedness parameter of the 
\mbox{NFPD} $\zeta$ equals the Bjorken variable
$x_{\rm Bj} = -q^2/2 \nu$, $\nu = P\cdot q$
\cite{MPW97}. This is clearly not the case for two produced particles.
Therefore it is natural to introduce the variable $x_p =
x_{\rm Bj}/\zeta$, \mbox{$0 \leq x_p \leq 1$} which describes the invariant
mass of the final $q\bar q$ state. Using furthermore the variables
\begin{equation}
Q^2=-q^2, \; 
y = \frac{P\cdot q}{P\cdot l}, \;
z_f = \frac{P\cdot p_f}{P\cdot q}, \;
t = (P-P')^2
\end{equation}
and $\varphi$ defined as the difference between the azimuthal angles of
$l$ and $p_f$ (see Fig.~\ref{kinem}) we get for the differential cross
section  
\begin{equation}
\frac{\d\sigma}
{\d x_{\rm Bj} \d y \d\zeta \d z_f \d t \d\varphi} =  
\frac{\alpha_{em}^2}{Q^4}\Bigg(\frac{1}{4\pi}\Bigg)^4 
 y L^{\mu \nu} H_{\mu \nu} \label{sig}
\end{equation}
where $L^{\mu\nu}$ is the usual leptonic tensor given by
\begin{eqnarray}
L^{\mu \nu} &=& \frac{1}{2}\sum_{\lambda '}
\bar u_\lambda (l)\gamma^\mu u_{\lambda '}(l')\bar u_{\lambda '}(l')
\gamma^\nu u_\lambda (l) \nonumber\\
&=& 2 l^\mu l^\nu - l^\mu q^\nu - l^\nu q^\mu -\frac{1}{2}Q^2
g^{\mu \nu} -\lambda \mathrm{i}
\varepsilon^{\mu \nu \varrho \sigma } l_\varrho q_\sigma \;.
\end{eqnarray}
Summing over the spins of the final nucleon 
we define analogously the hadronic tensor as
$\sum_{S'}H_{\mu\nu}=T^*_\mu T_\nu$ with 
\begin{eqnarray}
&&(2\pi)^4 \delta(p_f + p_f' - q - \zeta P) T_\mu \nonumber\\
&& = \int\d ^4 x\; \e^{-\mathrm{i} q\cdot x}\bra{q(p_f) \bar{q}(p_f') N(P')}
j_\mu(x) \ket{N(P)} \;.
\end{eqnarray}
$j_\mu(x)$ denotes the vector current that couples to the virtual
photon. In second order perturbation theory we may write
\begin{eqnarray}
&&(2\pi)^4 \delta(p_f + p_f' - q - \zeta P) T_\mu = 
\frac{\mathrm{i}^2 g^2}{2!}
\int \d^4 x_1 \int \d^4 x_2 \int \d^4 x_3 \; \e^{-\mathrm{i} q\cdot x_1}
\nonumber \\
&&\times\bra{q(p_f) \bar{q}(p_f') N(P')} 
T\{j_\mu(x_1) \, j^a_\alpha(x_2) \,j^b_\beta(x_3) A^{a\alpha}(x_2)
A^{b\beta}(x_3)\}
\ket{N(P)} 
\end{eqnarray}
with the color current $j^a_\alpha$ and the gluon field
$A^{a\alpha}$. We factorize quark and gluon fields in 
hard propagators and non perturbative matrix elements respectively
using the parameterization of nonforward matrix elements of bilocal
operators in terms of nonforward parton distributions \cite{Ra98}.
We project onto the twist-2 contributions by introducing 
a light-like vector $q_1^2 = (q + x_{\rm Bj} P)^2 = 0$ so that
we have for the quark matrix element
\begin{eqnarray}
&&\bra{N(P')} \bar{\psi}_f(x_2) \slas{q}_1\psi_f(x_1) \ket{N(P)} = 
\bar U (P') \slas{q}_1 U(P)
\nonumber \\
&&\times\int \d X
\Bigg[
\mathcal{F}^f_\zeta(X)
\e^{- \mathrm{i} X P\cdot x_1 + \mathrm{i} ( X-\zeta) P\cdot x_2}-
\mathcal{F}^{\bar f}_\zeta(X)   \e^{- \mathrm{i} X P\cdot x_2 + \mathrm{i}
( X-\zeta)
P\cdot x_1}\Bigg]\nonumber \\
&&+ \mathcal{K}{\rm -terms} \;,
\end{eqnarray}
\begin{eqnarray}
&&\bra{N(P')} \bar{\psi}_f(x_2) \gamma_5 \slas{q}_1\psi_f(x_1) \ket{N(P)} = 
\bar U (P') \gamma_5 \slas{q}_1 U(P)
\nonumber \\
&&\times\int \d X 
\Bigg[
\Delta \mathcal{F}^f_\zeta(X)
\e^{-\mathrm{i} X P\cdot x_1 + \mathrm{i} ( X-\zeta) P\cdot x_2}+
\Delta\mathcal{F}^{\bar f}_\zeta(X) \e^{- \mathrm{i} X P\cdot x_2
+ \mathrm{i} ( X-\zeta)
P\cdot x_1}\Bigg]\nonumber \\
&&+ \mathcal{K}{\rm -terms} \;,
\end{eqnarray}
and for the gluon fields
\begin{eqnarray}
&&\bra{N(P')} A^a_{\varrho}(x_2) A^a_{\sigma}(x_1) \ket{N(P)} \nonumber\\
&& = \frac{\bar U (P') \slas{q}_1 U(P)}{2 Pq_1} 
\left(
- g_{\varrho\sigma} + \frac{P_\varrho q_{1\; \sigma} + P_\sigma
q_{1\; \varrho}}{P q_1}\right) 
\nonumber \\
&&\times\int \frac{\d X}{2}
\frac{\mathcal{G}_\zeta(X)}{(X-\zeta + \mathrm{i}\varepsilon)
(X- \mathrm{i}\varepsilon)}
\Bigg[\e^{- \mathrm{i} X P\cdot x_1 + \mathrm{i} ( X-\zeta) P\cdot x_2} + 
      \e^{- \mathrm{i} X P\cdot x_2 + \mathrm{i} ( X-\zeta) P\cdot x_1}\Bigg]
\nonumber \\
&&+\frac{\bar U (P') \gamma_5 \slas{q}_1 U(P)}{2 Pq_1}
\mathrm{i} \varepsilon_{\varrho\sigma\lambda\eta}
\frac{P^\lambda q_1^\eta}{P \cdot q_1}
\nonumber \\
&&\times\int \frac{\d X}{2}
\frac{\Delta\mathcal{G}_\zeta(X)}{(X-\zeta + \mathrm{i}\varepsilon)
(X- \mathrm{i}\varepsilon)}
\Bigg[\e^{- \mathrm{i} X P\cdot x_1 + \mathrm{i} ( X-\zeta) P\cdot x_2} -
      \e^{- \mathrm{i} X P\cdot x_2 + \mathrm{i} ( X-\zeta) P\cdot x_1}\Bigg]
\nonumber \\
&&+ \mathcal{K}{\rm -terms} \;.
\end{eqnarray}
We have not written the gauge links explicitly. In the following we
will neglect the $\mathcal{K}$-terms that are proportional to the
momentum transfer to the nucleon.
Applying Fierz transformation, restricting ourselves to unpolarized
and longitudinal polarized nucleons and taking into account the
equations of motion we find
\begin{eqnarray}
T_\mu &=& T_\mu^{(g)}+T_\mu^{(q)}\;,\\
T_\mu^{(g)} &=& -\frac{T_F\;q_f}{N_C} g^2
\frac{\bar U(P') \slas{q}_1 U(P)}{2 Pq_1}
\frac{1}{\nu^2 z_f(1-z_f)}\nonumber \\
&& \times\int_0^1 \frac{\d X}{2}
\frac{\mathcal{G}_\zeta(X)}{(X-\zeta + \mathrm{i}\varepsilon)^2
(X - \mathrm{i}\varepsilon)^2}
\nonumber \\
&& \times \Bigg\{
\bar u (p_f) \Slas{P} v(p_f') 
\Big ( (X-\zeta)^2 + X^2 \Big ) P_\mu
\nonumber \\
&& \quad +
\bar u (p_f) \gamma_\mu v(p_f') 
\Big ( (X-\zeta)^2 + X^2-2 x_p \zeta^2
\Big ) \frac{\nu}{\zeta}
\Bigg\}
\nonumber \\
&& -\frac{T_F\;q_f}{N_C} g^2
\frac{\bar U(P') \gamma_5 \slas{q}_1 U(P)}{2 Pq_1}
\frac{1}{\nu^2 z_f(1-z_f)}\nonumber \\
&& \times\int_0^1 \frac{\d X}{2}
\frac{\Delta \mathcal{G}_\zeta(X)(2X - \zeta)}
{(X-\zeta + \mathrm{i}\varepsilon)^2 (X - \mathrm{i}\varepsilon)^2}
\nonumber \\
&& \times \Bigg\{
\bar u (p_f) \Slas{P} \gamma_5 v(p_f') 
\Big [2 p_{f\mu}- \zeta P_\mu \Big ] 
+ \bar u (p_f) \gamma_\mu \gamma_5 v(p_f') 
(1 - 2z_f) \nu
\Bigg\} \;,\label{gamp}\\
T_\mu^{(q)}
&=& -\frac{1}{4} \frac{C_F\;q_f}{N_C} g^2 \frac{\bar U(P') \slas{q}_1
U(P)}{2 Pq_1} \frac{1}{\nu^2 z_f(1 - z_f)}
\nonumber\\ && \times \int_0^1 \d X
\frac{[\mathcal{F}^f_\zeta(X) + \mathcal{F}^{\bar f}_\zeta(X)](2X - \zeta)}
{(X-\zeta + \mathrm{i}\varepsilon) (X - \mathrm{i}\varepsilon)
(X-\zeta x_p + \mathrm{i}\varepsilon)
(X-\zeta (1-x_p) - \mathrm{i}\varepsilon)}
\nonumber \\ && \times
\Bigg\{
\bar u (p_f) \Slas{P}  v(p_f')
\Big[\Big(X(X-\zeta) + \zeta^2 x_p (1 - z_f)\Big)P_{\mu}
+\zeta x_p (2z_f - 1)p_{f\mu}\Big]
\nonumber \\ && \quad
+\bar u (p_f) \gamma_\mu v(p_f')
\frac{\nu}{\zeta}
\Big(X(X - \zeta)-\zeta^2 x_p(x_p-2z_f(1 - z_f))\Big)\Bigg\}
\nonumber \\
&& -\frac{1}{4} \frac{C_F\;q_f}{N_C} g^2 \frac{\bar U(P')
\slas{q}_1 U(P)}{2 Pq_1} \frac{1}{\nu^2 z_f(1 - z_f)}
\nonumber\\ && \times \int_0^1 \d X
\frac{\mathcal{F}^f_\zeta(X) - \mathcal{F}^{\bar f}_\zeta(X)}
{(X-\zeta + \mathrm{i}\varepsilon) (X - \mathrm{i}\varepsilon)
(X-\zeta x_p + \mathrm{i}\varepsilon)
(X-\zeta (1-x_p) - \mathrm{i}\varepsilon)} 
\nonumber \\ && \times
\Bigg\{
\bar u (p_f) \Slas{P}  v(p_f')
\Big[\Big(\zeta X (\zeta - X)(1+x_p(2-4z_f))-\zeta^3 x_p (1 - z_f)
\Big)P_{\mu} 
\nonumber \\ && \qquad
+ \Big(2X(X - \zeta)+\zeta^2 x_p\Big)p_{f\mu}\Big]
\nonumber \\ && \quad        
+\bar u (p_f) \gamma_\mu v(p_f')
\nu \Big(X(\zeta - X)-\zeta^2 x_p^2\Big)(2z_f - 1)\Bigg\}\nonumber\\
&&-\bigg\{\slas{q}_1\rightarrow\gamma_5\slas{q}_1,
\Slas{P}\rightarrow \Slas{P}\gamma_5,
\gamma_\mu\rightarrow\gamma_\mu\gamma_5,
\mathcal{F}^f\rightarrow\Delta\mathcal{F}^f,
\mathcal{F}^{\bar f}\rightarrow-\Delta\mathcal{F}^{\bar f} \bigg\}\;.
\nonumber\\
\end{eqnarray}
A typical diagram contributing to $T_\mu^{(g)}$
is shown in Fig.~\ref{dijet} on the left, an example for a diagram
contributing to $T_\mu^{(q)}$ in the same figure on the right.
The color factors are denoted in the common way $C_F = 4/3$, $T_F = 1/2$,
and the number of colors $N_C = 3$. $q_f$ is the electric charge of the
produced quark.
It is easy to see that by going to the limit of a massless final state, 
i.e.~$x_p = 1$, and contracting the amplitude with the  
polarization vector of a longitudinal polarized photon the results for 
exclusive meson production are reproduced. See for instance 
\cite{MPW97}.
In such a kinematical limit $z_f$ corresponds to the light 
cone momentum fraction carried by the produced quark. The appearing soft
divergences $1/[z_f(1-z_f)]$ then do not pose any problems for 
meson wave functions that vanish at the end points. In our case we
have to stay off these divergences using cuts for $z_f$. Such cuts
are demanded by the acceptance of the meson detectors because of the
relation $z_f \approx z_{\pi}:= Pp_{\pi}/Pq$.
Note that upon contraction with a longitudinal polarization vector and
the limit $x_p = 1$ all dependence on the polarized gluon distribution
$\Delta\mathcal{G}$ cancels out.

Contracting $T_\mu^*T_\nu$ with the leptonic tensor $L^{\mu\nu}$ leads
us to the differential cross
section $\frac{\d\sigma}{\d x_{\rm Bj}\d y\d\zeta\d z_f\d t\d\varphi}$ in
terms of the nonforward distributions. 
To the polarized cross section only the interference term of  
polarized parton densities and unpolarized densities contribute. All other
combinations contribute to unpolarized scattering.
For the phenomenological analysis we have
to perform the remaining integrations over $X$ in
the amplitudes. To do so we need some model for the nonforward parton
distributions.

\section{Nonforward Parton Distributions}

The nonforward parton distributions used in the calculations of the
previous section are given by an integral over the corresponding double
distributions
\begin{equation}
\mathcal{F}_\zeta(X) = \int_0^a 
{F(X-\zeta y,y)} \d y \label{nonf}
\end{equation}
with $a=\mbox{min}\{\frac{X}{\zeta},\frac{(1-X)}{(1-\zeta)}\}$ \cite{Ra98}.
If we write a double distribution as a product of the corresponding usual
forward distribution and a profile function
\begin{equation}
F(x,y)=f(x)\pi(x,y)
\end{equation}
the latter has to fulfill the conditions
\begin{eqnarray}
\pi (x,y)=\pi (x,1-x-y) \;, \qquad
\int_0^{1-x}\pi (x,y) \d y = 1\;.
\end{eqnarray}
The symmetry $y \leftrightarrow 1-x-y$ was noted in \cite{MPW97}.
As a model for the profile functions we take
$\pi(x,y)={\frac{6 y(1-x-y)}{(1-x)^3}}$ 
for the quark distributions and
$\pi(x,y)={\frac{30 y^2(1-x-y)^2}{(1-x)^5}}$
for the gluon distributions \cite{Ra98}.
For the (forward) parton distributions we use the (slightly simplified)
NLO parameterizations of \cite{MRSA} in the unpolarized and \cite{GS95} in
the polarized case at the scale $Q_0^2=4 \mbox{GeV}^2$.
From these models we get the nonforward parton distributions using
(\ref{nonf}). In the following we will neglect all effects of $Q^2$ 
evolution of the distribution functions.
Special care is
needed for the quark sea distribution for which the original
parton distribution is more singular than $1/x_{\rm Bj}$. In this case
the integral (\ref{nonf}) is not defined for $X \leq \zeta$. The
problem can be solved using the anti-symmetry of the remaining part of
the corresponding integrand:
\begin{eqnarray}
\int_0^1 \mathcal{A}(\zeta,X)\mathcal{F}_\zeta(X) \d X
&=& \int_{-1+\zeta}^\zeta \mathcal{A}(\zeta,\zeta -X)\mathcal{F}_\zeta
(\zeta -X) \d X \nonumber \\
&=& \frac{1}{2}\int_{-1+\zeta}^1 \mathcal{A}(\zeta,X)\nonumber\\
&&\times
\Big(\mathcal{F}_\zeta(X)\Theta(X)
-\mathcal{F}_\zeta(\zeta -X)\Theta(\zeta -X) \Big) \d X \nonumber \\
&=& \int_{\zeta /2}^1 \mathcal{A}(\zeta,X)\nonumber\\
&&\times\Big(\mathcal{F}_\zeta(X)-\mathcal{F}_\zeta(\zeta -X)\Theta(\zeta -X)
\Big)\d X\;.
\end{eqnarray}
Here the divergences cancel in the difference
$\mathcal{F}_\zeta(X)-\mathcal{F}_\zeta(\zeta -X)$.

In order to determine the cross section integrated over $t$ we assume
that the $t$-dependent nonforward distributions factorize into a
nucleon form factor and the nonforward distribution at $t=0$
\begin{equation}
\label{formfaktor}
\mathcal{F}_\zeta(X,t)=\mathcal{F}_\zeta(X,0)F(t)=\mathcal{F}_\zeta(X)F(t) \;.
\end{equation}
This behavior is suggested (at least for small values of $t$) by the
condition $\int_0^1 \mathcal{F}_\zeta(X) \d X = F(t)$
\cite{Ra97} (see also \cite{MPW97}).
As models for the form factors we take
\begin{equation}
F(t)=\Bigg(\frac{1}{1-t/\Lambda^2}\Bigg)^\alpha
\end{equation}
with $\alpha=2$, $\Lambda=0.84\;{\rm GeV}$ in the case of quark
distributions and $\alpha=3$, $\Lambda=2.8\;{\rm GeV}$ in
the case of gluon distributions. The quark form factor
is taken from measurements of elastic electron nucleon scattering
\cite{Weise}, the gluon form factor from QCD sum rule calculations
\cite{BGMS93}. Since these form factors
decrease rapidly for increasing $|t|$ the integrals over $t$ will be
dominated by small $|t|$. Therefore it is justified to
keep $t=0$ in the other contributions to the differential cross
sections. In this approximation the $t$-integration can be carried out
and leads to factors of the form $\int_{-\infty}^0 F_1(t)F_2(t) \d t$
where $F_i$ are nucleon form factors for quark or gluon fields
respectively.

\section{Fragmentation into Mesons}

\begin{figure}
\centerline{\psfig{figure=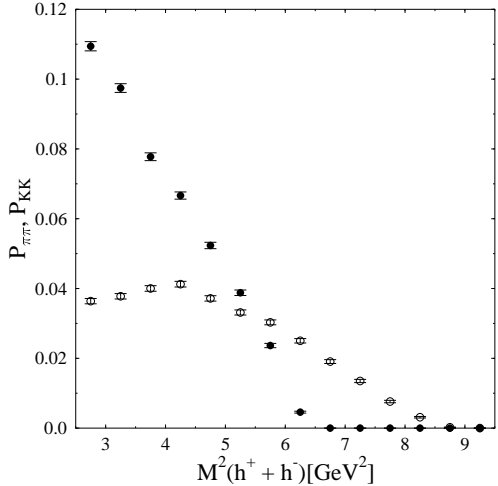,width=8cm}}
\caption{Probability of creating a $\pi^+\pi^-$ pair (solid circles) or
a $K^+K^-$ (open circles) pair in a quark-anti quark fragmentation
process with center of mass energy squared $M^2(h^+ + h^-)$.}
\label{plot1}
\end{figure}

To describe the fragmentation process of
the two outgoing quarks into a meson pair we use JETSET which is based
on the LUND string fragmentation model \cite{strgfrg,Sj95}.
In Fig.~\ref{plot1} we plot the probability to produce a pion or kaon
pair from two quarks as a function of the cm-energy 
$M^2(h^+ + h^-) = (p_f + p_f')^2 = (\zeta P +q)^2 = 
2 \nu \zeta(1 - x_p)$.
For this plot we assumed that all quark-flavors are produced in equal
ratios as soon as $M^2(h^+ + h^-)$ exceeds the respective
mass threshold.

In order to evaluate the cross section for $\pi^+\pi^-$ and
$K^+ K^-$ pair production in lepton proton scattering
the events generated by JETSET are weighted with our
differential cross section for quark pair production $\frac{\d\sigma}
{\d x_{\rm Bj}\d y\d\zeta\d z_f\d\varphi}$.
Fig.~\ref{plot1} shows that the center of mass energy squared
should be less than $8\;{\rm GeV^2}$ to give sizable meson pair production.
In order to avoid regions where we cannot trust the program JETSET
any longer we further have to avoid too small values of $M^2(h^+ + h^-)$.
For our simulations we work in the proton rest frame with a lepton energy of
$27.5\;{\rm GeV}$ and use an
acceptance slightly smaller than the HERMES acceptance. We require the
angles of all detected particles (electrons, pions and kaons) with respect to
the electron beam axis to lie
in the range $ 0.04 < \tan \theta < 0.140 $. 
Furthermore, the detected electron should have an energy exceeding
$1\;{\rm GeV}$ and the detected mesons an energy larger than 
$4\;{\rm GeV}$. We also require both mesons to have a minimum
transverse momentum in the laboratory frame of $0.5\;{\rm GeV}$.
Further we imposed the cut $p_T^2 > 1\;{\rm GeV^2}$ (with $p_T$ being the
transverse momentum in the center of mass frame of the photon and the
incoming nucleon) in order to keep the relevant scale
$\mu^2=p_T^2(1+\frac{Q^2}{M^2(h^+ + h^-)})$ sufficiently large.
For the remaining variables we used the cuts
$0.02 < x_{\rm Bj} < 0.05$,
$0.29 < y  < 0.96$,
$0.1 < \zeta < 0.3$, and
$0.04 < z_f < 0.96$
where the lower cut in $\zeta$ was dictated by numerical reasons.
The cuts in $z_f$ are part of the
$z-s$ jet scheme which we applied in our simulations. It is also
used in the standard Monte Carlo leptoproduction programs \cite{ingelman}.

\section{Results and Discussion}

In Table~\ref{tab} we have given the various 
contributions to the total
cross section for meson pair production.
\begin{table}
\caption{The different contributions to the total cross sections.}
\begin{tabular}{ccc}
\\
\hline
& cross sect. $\pi^+\pi^-$ (pb) &  cross sect. $K^+K^-$ (pb) \\
\hline
\hline
$\sigma_{gg}$                   & $ \;\;3.2\;10^{-1}$ & $ \;\;2.8\;10^{-1}$ \\
$\sigma_{\Delta g\Delta g}$     & $ \;\;4.2\;10^{-2}$ & $ \;\;4.1\;10^{-2}$ \\
$\sigma_{g\Delta g}$            & $ \;\;6.3\;10^{-2}$ & $ \;\;5.3\;10^{-2}$ \\
\hline
$\sigma_{qq}$                   & $ \;\;5.4\;10^{-2}$ & $ \;\;4.0\;10^{-2}$ \\
$\sigma_{\Delta q\Delta q}$     & $ \;\;3.3\;10^{-3}$ & $ \;\;2.9\;10^{-3}$ \\
$\sigma_{q\Delta q}$            & $ \;\;7.4\;10^{-3}$ & $ \;\;4.8\;10^{-3}$ \\
\hline
$\sigma_{qg}$                   & $    -1.2\;10^{-1}$ & $    -8.8\;10^{-2}$ \\
$\sigma_{\Delta q\Delta g}$     & $    -4.6\;10^{-3}$ & $    -7.7\;10^{-4}$ \\
$\sigma_{q\Delta g+ g\Delta q}$ & $    -4.3\;10^{-2}$ & $    -2.9\;10^{-2}$ \\
\hline
total spin average              & $ \;\;2.9\;10^{-1}$ & $ \;\;2.8\;10^{-1}$ \\
total spin asymmetry            & $ \;\;2.8\;10^{-2}$ & $ \;\;2.9\;10^{-2}$ \\
\hline\\
\end{tabular}
\label{tab}
\end{table}
Contributions from the
polarized and unpolarized gluonic \mbox{NFPD} are denoted by the indices
$g$ and $\Delta g$,
contributions from the quark \mbox{NFPDs} by $q$ and $\Delta q$ respectively.
Since the cross section is quadratic in the parton distributions
we find pure gluonic contributions of the form 
$\sigma_{gg}$, $\sigma_{\Delta g \Delta g}$, and $\sigma_{g \Delta g}$,
similar expressions for the pure quarkonic part, and interference
contributions $\sigma_{qg}$, $\sigma_{\Delta q \Delta g}$, and
$\sigma_{q \Delta g + g \Delta q}$.
Only the interference terms of polarized and unpolarized parton
densities $\sigma_{g \Delta g}$, $\sigma_{g \Delta g}$, and
$\sigma_{g \Delta q + q \Delta g}$
depend on the alignment of the spins of the incoming electron and proton and
thus contribute to the spin asymmetries
$\Delta\sigma:=\frac{1}{2}
(\sigma_{\uparrow\downarrow}-\sigma_{\uparrow\uparrow})$.
($\sigma_{\uparrow\downarrow}$ is the cross section for antiparallel
lepton spin and nucleon spin.)
Comparing the contributions to the spin average we see that
$\sigma_{\Delta g \Delta g}$, $\sigma_{\Delta q \Delta q}$, and
$\sigma_{\Delta q \Delta g}$ are roughly one order of magnitude
smaller than $\sigma_{gg}$, $\sigma_{qq}$ and $\sigma_{qg}$ respectively.

For our \mbox{NFPD} Models, the gluon contribution dominates the cross section,
both in the unpolarized and polarized channel.
Note that the normalization of the cross section
depends crucially on the form factors (\ref{formfaktor}).
For the used input distributions we finally get 
a spin asymmetry of order 10\%.

\begin{figure}
\centerline{\epsfig{figure=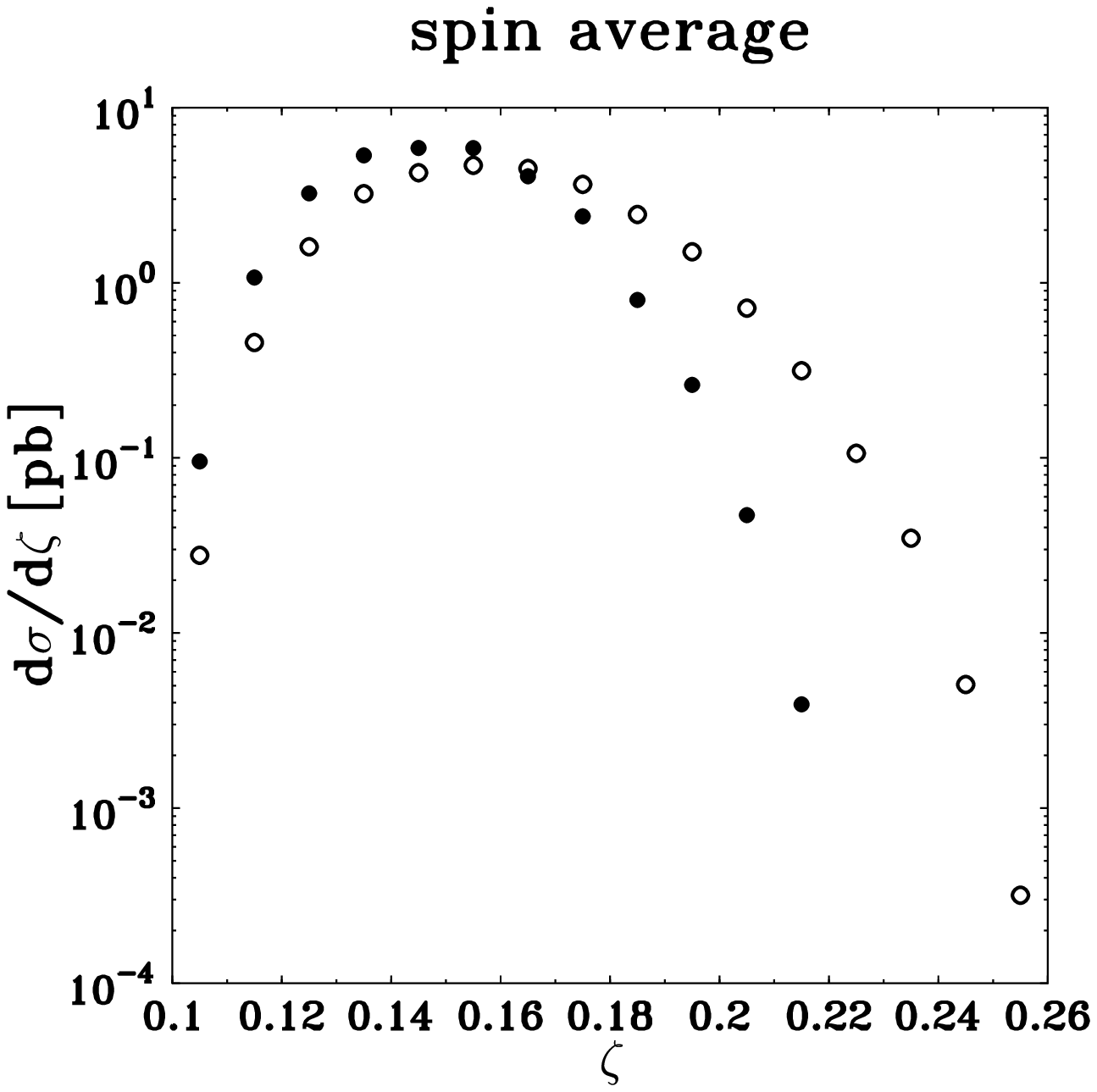,width=6.2cm}
            \epsfig{figure=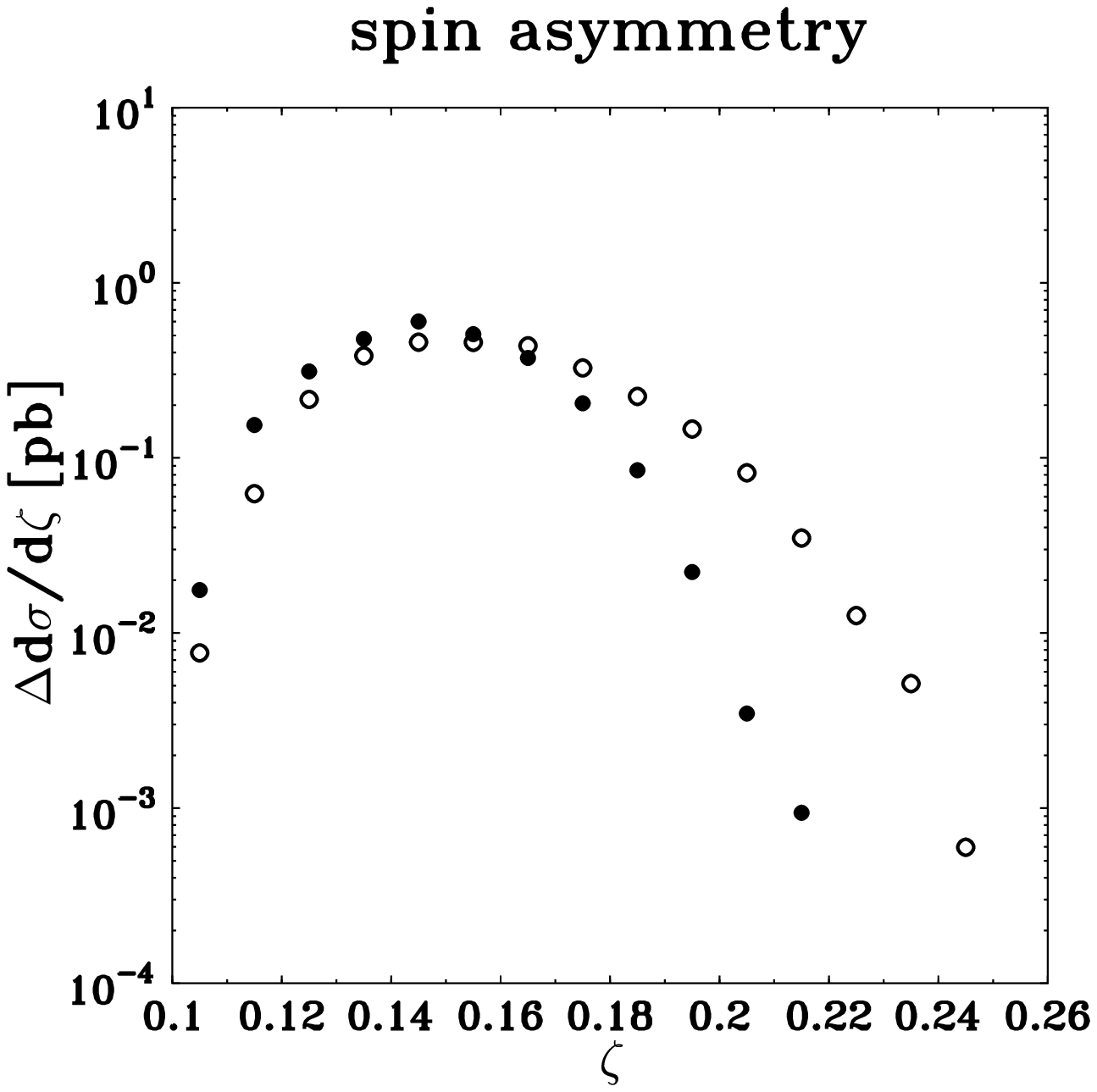,width=6.2cm}}
\vspace{-0.4cm}
\centerline{\epsfig{figure=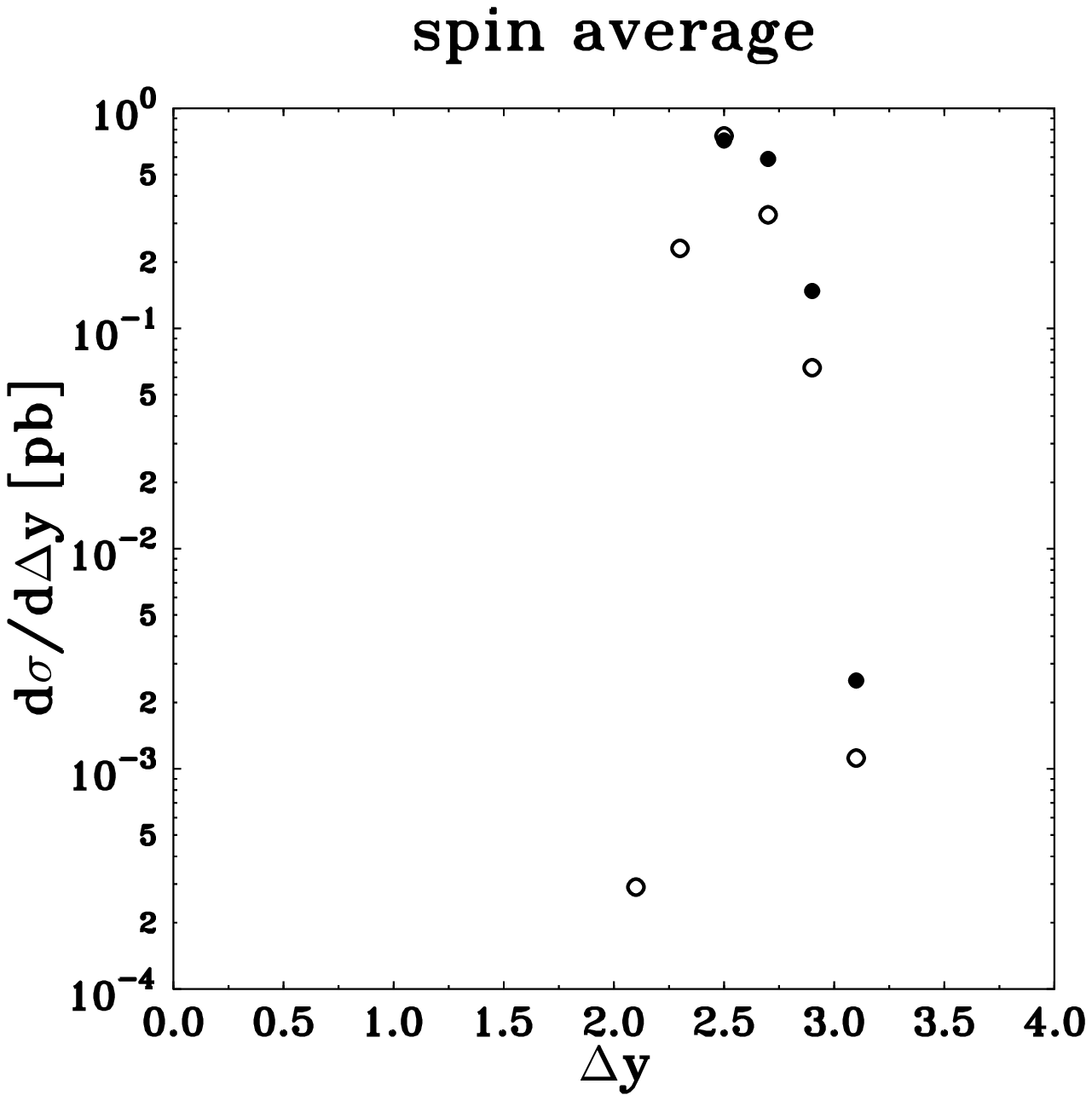,width=6.2cm}
            \epsfig{figure=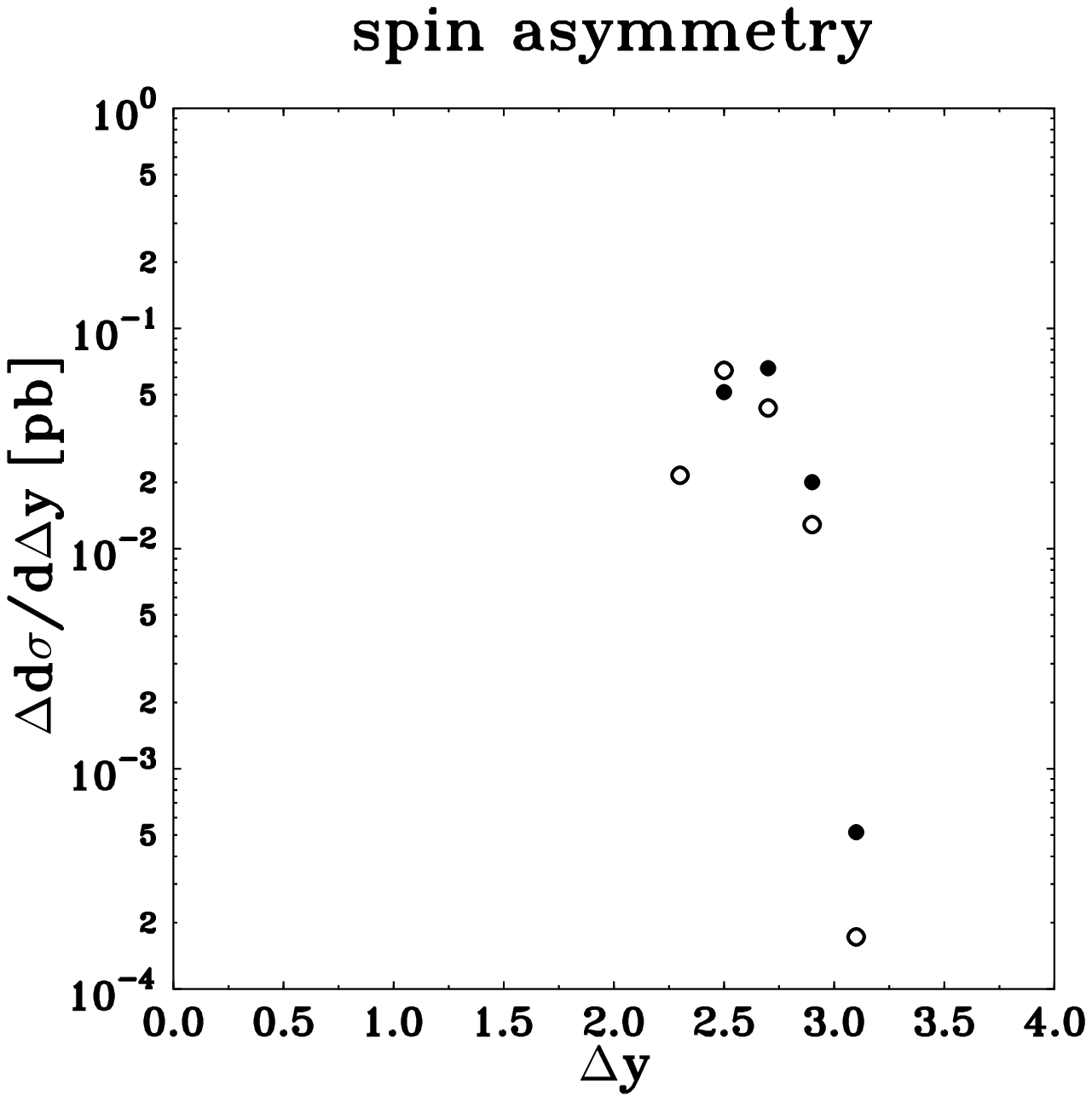,width=6.2cm}}
\vspace{-0.4cm}
\centerline{\epsfig{figure=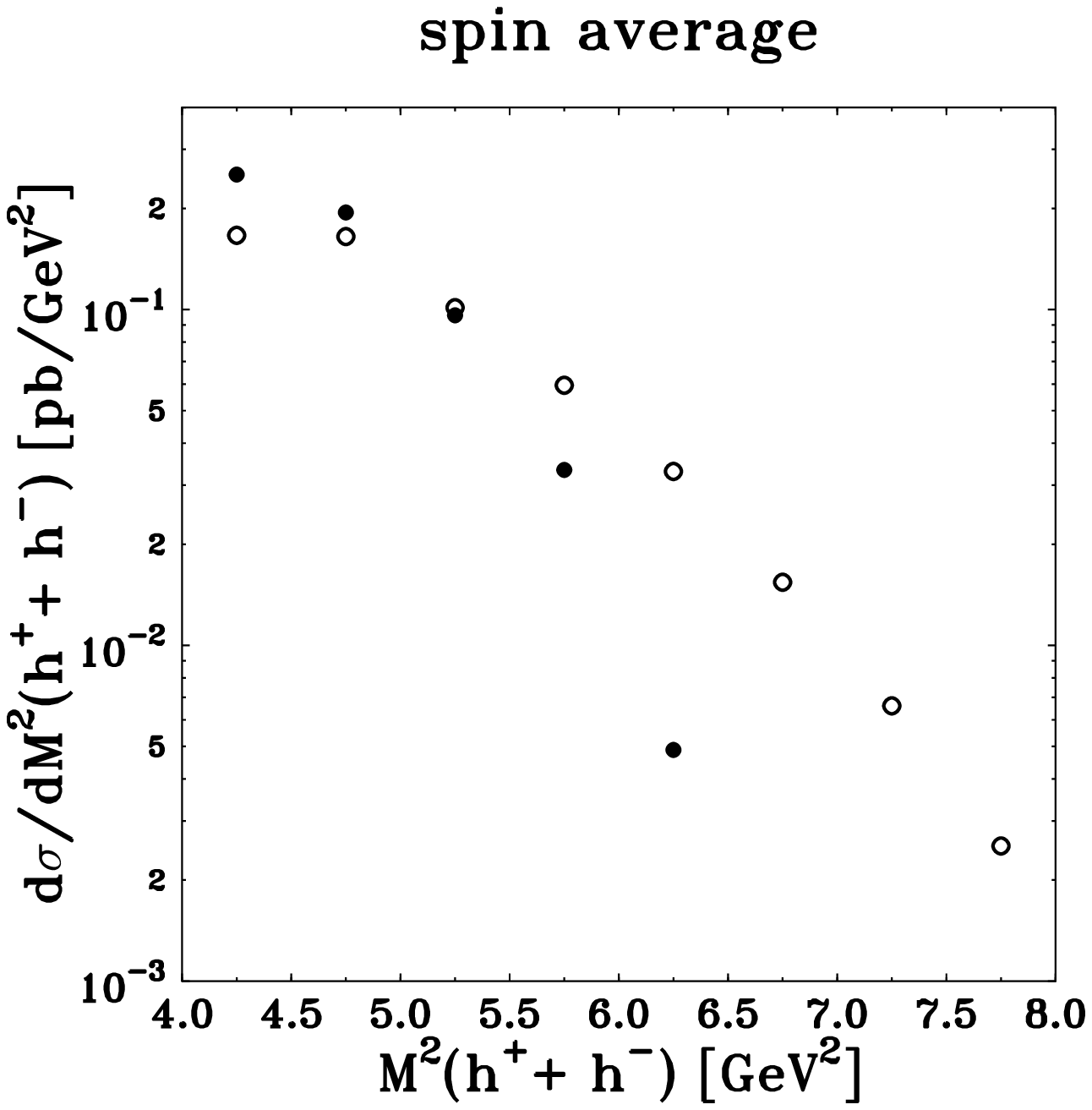,width=6.2cm}
            \epsfig{figure=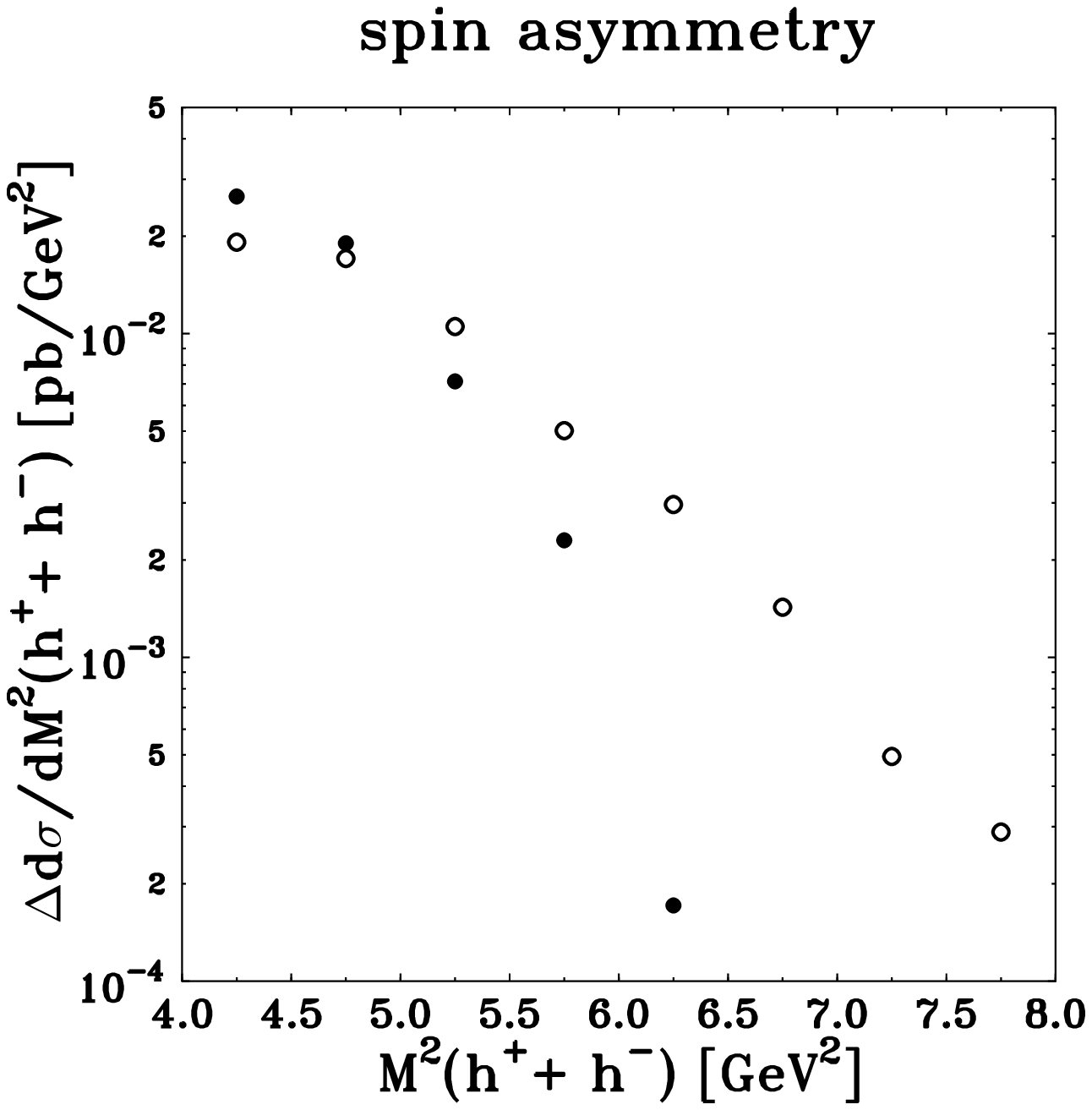,width=6.2cm}}
\caption{The unpolarized (on the left) and polarized (on the right)
differential cross section for diffractive $\pi^+\pi^-$ (solid
circles) and $K^+K^-$ (open circles) production with respect to the
variables $\zeta$, $\Delta y :=
\min\{y_{\rm particle \;1},y_{\rm particle\; 2}\}
- y_{\rm \;outg.\; proton}$, and $M^2(h^+ +h^-)$.}
\label{cross}
\end{figure}

The unpolarized and polarized cross sections
are plotted in the upper pictures in Fig.~\ref{cross}
as functions of the skewedness $\zeta$.
The cross sections receive their
dominating contribution from the small values of $\zeta$. In our
analysis we restrict ourselves on skewedness values larger than
$\zeta > 0.1$.
To go to smaller values of $\zeta$ ultimately would lead
to the requirement that $Q^2 \to 0$ and thus would lead to 
the region of photoproduction. In that kinematical domain the scattered
electron is not detectable and thus an unambiguous reconstruction of the
final hadronic state is not possible unless the outgoing proton could
be detected by a recoil spectrometer.
One method to distinguish the elastic events from inelastic ones in
this case is provided by the existence of a rapidity gap. The plots in
the middle of Fig.~\ref{cross} show that the difference between the
minimum of the meson rapidities  and the rapidity of the outgoing
proton $\Delta y$ is for all diffractive events larger than $2$.
In the lower diagrams in Fig.~\ref{cross} we plotted the cross section
versus the invariant mass squared of the produced particles. These
plots show the increase of the ratio between the kaon
and pion rates with the invariant mass $M(h^+ + h^-)$ as suggested by
Fig.~\ref{plot1}.
The shape of the polarized and unpolarized cross sections
is similar.

We have analyzed diffractive $\pi^+\pi^-$ and $K^+K^-$ production as a
tool for the determination of polarized and unpolarized
nonforward parton distributions. We implemented the realistic cuts for
the HERMES experiment and required sufficiently large $p_T$ to justify
our leading order treatment. With these cuts the cross sections are
too small to make this channel a promising one for HERMES. Successfull
measurements would probably require an ENC-type machine \cite{GSI}.

\begin{ack}
The authors thank M.~Diehl, P.~Hoyer, L.~Mankiewicz, G.~Piller, M.~Polyakov, 
A.~Radyushkin, M.~Strikman, and O.~Teryaev for useful discussions.
This work has been supported by DFG (Graduiertenkolleg), Studienstiftung, and
BMBF. M.M. thanks the Nordic Insitute for Theoretical Physics
for financial support and hospitality.
\end{ack}

\end{document}